\documentclass{jfm}
\usepackage{graphicx}
\usepackage{epstopdf, epsfig}
\usepackage[dvipsnames]{xcolor}

\definecolor{mlcolor}{RGB}{0, 0, 0}
\newcommand{\ml}[1]{\textcolor{mlcolor}{{#1}}}
\newcommand{\mkl}[1]{\textcolor{black}{{#1}}}
\definecolor{rwgreen}{RGB}{0, 0, 0}
\definecolor{pspink}{RGB}{0, 0, 0}
\definecolor{blu}{RGB}{0, 0, 0}
\newcommand{\rw}[1]{\textcolor{black}{#1}}

\graphicspath{{./figures/}}
\usepackage{mathtools,mathrsfs,lipsum,color}
\newcommand{\pd}{\partial}
\newcommand{\mvec}[1]{\boldsymbol{#1}}

\title{Stability of sedimenting flexible loops}
\shorttitle{Sedimenting flexible loops}
\shortauthor{R. Waszkiewicz, P. Szymczak and M. Lisicki}

\author{Radost Waszkiewicz\corresp{\email{rwaszkiewicz@fuw.edu.pl}},
	Piotr Szymczak,
 \and Maciej Lisicki\corresp{\email{mklis@fuw.edu.pl}}}

\affiliation{Institute of Theoretical Physics, Faculty of Physics, University of Warsaw, Pasteura 5, 02-093 Warsaw, Poland}

\begin{document}

\maketitle

\begin{abstract}
We study the behaviour of circular flexible loops sedimenting in a viscous fluid by numerical simulations and linear stability analysis. The numerical model involves a local slender-body theory approximation for the flow coupled to the Euler-Bernoulli elastic forces for an inextensible fibre. Starting from an inclined circle, we simulate the dynamics using truncated Fourier modes to observe three distinct regimes of motion: absolute stability, two-, and three-dimensional dynamics, depending on the relative importance of elastic and gravitational forces. We identify the governing parameter and develop a simple semi-analytic stability criterion, which we verify numerically. In all cases, sedimenting loops converge to stable, planar \rw{shape} equilibria with one free parameter related to the initial conditions and material properties of the fibre.


\end{abstract}


\section{Overview}
Biological processes are one of many inspirations of elastohydromechanics \citep{Schoeller:2019,Shelley:2000}. Slender biological objects emerge in multiple contexts, motivating detailed investigation. 
\rw{Starting from the sub-cellular level, examples include DNA and protein folding dynamics \citep{Goldstein1995}, lipids usually forming cell walls assembling into long filaments \citep{Rudolph1991}, or microtubules helping healing by contracting wounds \citep{Ehrlich1977}. Another large area of interest is motility -- cells moving inside a fluid environment or cells inducing motion of a fluid. An iconic example of such motion are sperm cells. On closer investigation it turns out that flexibility plays an important role in their motion, and the interplay between elasticity and viscous forces causes changes of the beating pattern in response to the changing environment \citep{Gaffney2011,Lagomarsino2003,Fauci2006}. Biological 'optimisation' for viscosity gradients can also be found in mucus transport inside the lungs where the correct length, stiffness and active deformation of cilia provides the necessary movement of multiple layers of fluid with varying viscosity, essential for healthy respiration \citep{Fulford1986}. On larger length scales, bacterial complexes were observed joining into elongated structures exhibiting complex dynamics because of elastohydrodynamic effects \citep{Goldstein1998,Mendelson1995}.}
Most of these examples are set in a microscale context, and thus the observed dynamics are dominated by the viscous interactions with the surrounding fluid \citep{Lauga:2009}. 

Flexible fibres with free ends have been studied in multiple settings including their sedimentation, both experimentally \citep{Herzhaft:1999} and numerically \citep{Lei:2013}. The free-end configuration was investigated first, because methods of producing slender filaments were already developed, and because it is of interest for both industrial applications and biological settings. Further, one-dimensional structure provides a particularly elegant, treatable, and successful way of modelling \citep{Wiggins1998}. 

In the case of low Reynolds number flows in such settings, elastic elongated filaments can be modelled using various simulation methods, for example: immersed boundary (IB) method \citep{Peskin:2002}, regularised singularity methods \citep{Cortez:2005}, bead-spring models {\citep{slowicka2015,Kuei_2015,Schoeller:2019}}, and discrete and continuous variants of slender-body theory (SBT) \citep{Saintillan:2005,Tornberg:2004}. \ml{The reduced dimensionality of the filament offers a computational advantage, which has been used in variants of the IB technique to study the whirling instability of spinning filaments \citep{limpeskin2004whirling} and hydrodynamic bundling of bacterial flagella \citep{limpeskin2012bundling}. A combination of SBT and regularised Stokeslet method has also been formulated by \citet{cortez2012slender} and profitably applied e.g. to explain the motion of flagella in dinoflagellates \citep{nguyen_ortiz_cortez_fauci_2011}. 
See \citet{nguyen2014amm} for a review of this approach.} On the other hand, in methods which treat the filament as fundamentally one-dimensional, such as SBT, one faces problems when the mesh along the filament is too fine (small in comparison to the reduced length scale), even when smoothing the integral kernels, as discussed by \citet{Tornberg:2004}.
Finally, any numerical scheme for the time evolution of elastic filaments must address the stiffness of the equations caused by presence of high-order spatial derivatives in the equations of motion responsible for bending rigidity. Due to the very high rate at which disturbances of small wavelength are damped, the issue of stiffness becomes even more pronounced with finer mesh sizes.
Practically all numerical works to date use an implicit integration scheme, while here we present a different approach.

We focus on a different configuration -- looped filaments with no free ends. \ml{The dynamics of microscale loops in viscous flows was previously analysed in the context of growing smectic-A liquid crystals, which were modelled by \citet{Shelley:2000} using SBT, and for circular filaments with a non-zero inherent twist and bend, explored using a variant of the IB method by \citet{lim2008dynamics} in the context of over- and underwinding of DNA leading to dynamic transitions of shapes.} Our work is motivated by the experimental work of \citet{Alizadehheidari:2015} on circular DNA confined to nanofluidic channels (and in particular its breaking), and \citet{Koche:2020} linking extrachromosomal circular DNA properties with neuroblastoma, {and partly inspired by previous numerical work using bead-spring hydrodynamic models \citep{Gruziel:2019}}. Electrophoretic and ultracentrifugation measurements of mobility pose questions about what constitutes a flexible regime, correct values of drag coefficients, and stress distribution along the filament. %

The inclusion of elasticity is necessary to analyse the aforementioned systems. It was observed experimentally for red blood cells \citep{Jay:1972} and numerically for flexible chains of beads \citep{Gruziel:2018,Gruziel:2019} that high flexibility leads to a change in orientation (and sometimes shape) of sedimenting objects, affecting their sedimentation speed. Independently, in the case of linear filaments, \citet{Reichert:2009} observed that including elasticity can change the behaviour qualitatively when looking at bundling vs. non-bundling flagella. Even for a fixed shape, a change in orientation can alter the sedimentation velocity by $25\%$, as shown by \citet{Tchen:1954} and in further analytical solutions for sedimenting tori \citep{Cox:1970,johnson_wu_1979,majumdar1977}.  Furthermore, as noted by \citet{Box:2018} and \citet{Kodio:2019}, dynamical buckling can occur in similar settings, resulting in significant shape changes of the filaments. In bead-spring models, \citet{Gruziel:2019} observed the existence of an elasticity threshold beyond which flexible loops undergo significant changes in sedimentation dynamics. While stiff loops were seen to attain almost planar, oval shapes and sediment vertically or at an acute angle to gravity, depending on their stiffness, more flexible fibres exhibited a complex shape evolution. Our work aims to explore this stability threshold in slender-body dynamics, both analytically and in terms of numerical simulations.

In this work, we analyse the dynamics of slender elastic loops by linear stability analysis in a coupled elastohydrodynamic model, and by numerical simulations introducing a new method based on Fourier expansions. \rw{The mathematical elegance of the periodic boundary conditions allows us to simplify the theoretical analysis and gain an analytical insight into the stability question.} Our results contribute to the explanation of horizontal sedimentation preference. \rw{We also derive explicit expressions for tension distribution along the filament, which comply with the work of  \citet{Alizadehheidari:2015} on DNA loops in microfluidic channels, where typical locations of ruptures correspond to the highest tension in our model.}

\section{Qualitative description}

We focus on a thin, inextensible, looped elastic filament, settling in a viscous fluid under gravity. The filament has a length $L$ and bending stiffness $A$. We consider the dynamics in the Stokesian regime of low Reynolds numbers, where the fluid drag forces are proportional to the local velocity of the filament. Solutions for the terminal velocity for a rigid loop were known even before the development of the slender-body theory \citep{Tchen:1954,majumdar1977} and were tested experimentally in some cases \citep{amarakoon1982drag}. \rw{The distribution of shear forces from the fluid acting on toroidal particles in these solutions is not uniform and thus has to be balanced by an equal and opposite force from the particle. For loops that are not perfectly stiff, these forces may partially arise from elastic deformation.}

The presence of elastic forces can give rise to a complex dynamics of the sedimenting loop. To understand the stability of sedimenting circular loops, we first consider a simpler case. A classical example of beam instability under external compression has a solution known since the mid-eighteenth century \citep{Euler:1757}. Buckling under internal forcing (heavy column buckling under its own weight) was revisited later by \citet{Greenhill:1881}. For a beam of length $L$ under the action of an external force $F$, stability results from a competition of this forcing with the stiffness of the beam. Because the bending rigidity is quantified by the product $A=EI$, with $E$ being the Young's modulus of the beam material, and $I$ being its cross-sectional moment of inertia, the relevant dimensionless quantity is $EI/FL^2$, \rw{capturing the stiffness to external force ratio, and there exists a critical threshold value of this quantity above which the initial shape becomes unstable}.

We thus expect that a large enough compression is sufficient to destabilise an elastic filament. On the other hand, when we compute forces on a straight beam bent into the shape of a circle, we find that bending forces are balanced by the tension (negative compression, or stretching) in the beam of the value $T = EI / R^2$\rw{, where $R$ is the radius of the circle}. Hence, in the absence of the fluid, \rw{we have two mechanisms which stabilise the shape against perturbations}: the negative compression rate in the beam and \rw{the forces that resist bending}. The presence of a fluidic environment introduces additional hydrodynamic forces that may be responsible for a local compression of the sedimenting loop.

\begin{figure}
    \centering
    \includegraphics[width=0.6\linewidth]{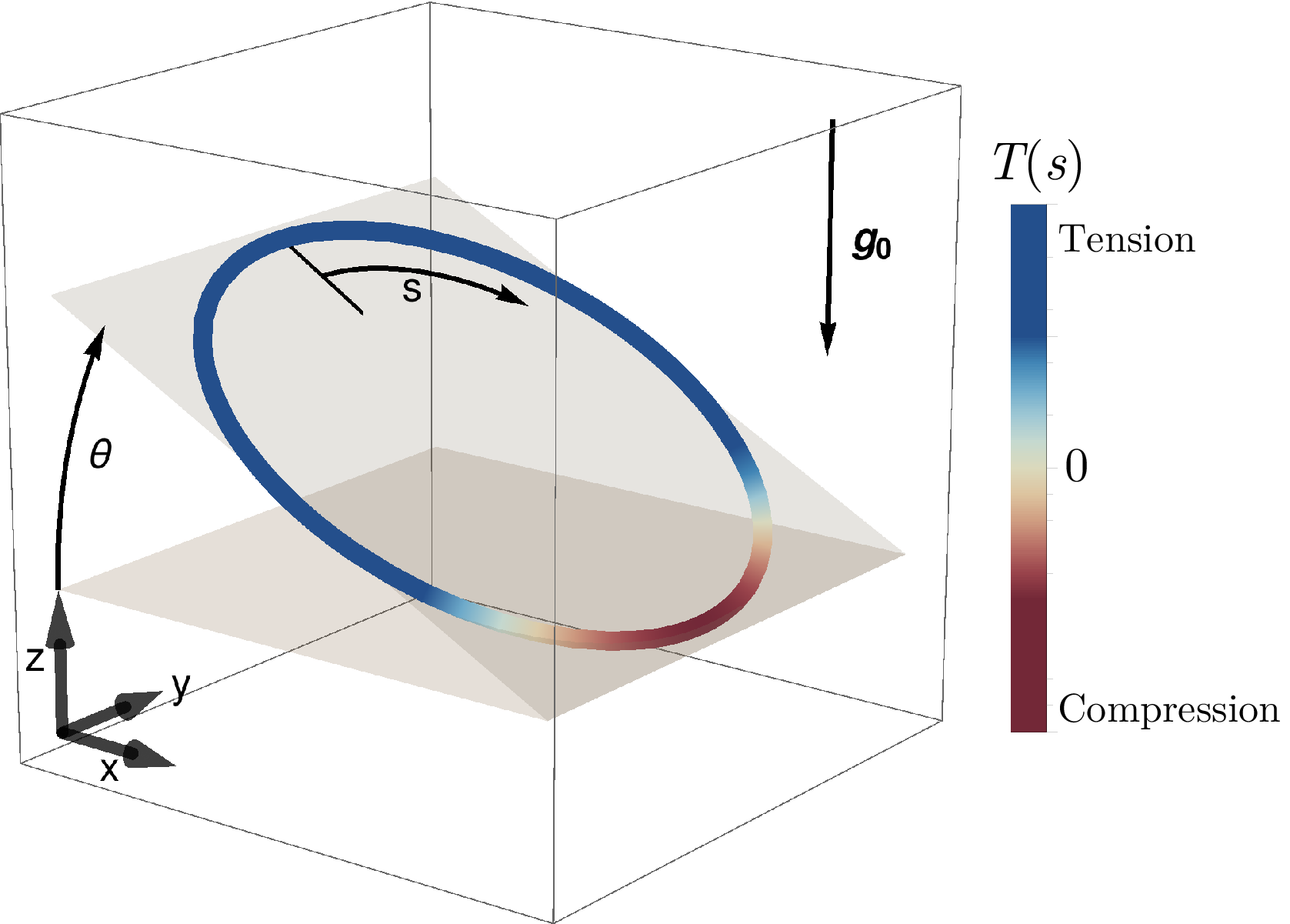}
    \caption{\rw{A 3D sketch of the studied system. The direction of gravity $\mvec{g_0}$, angle of inclination $\theta$ and the arc length variable $s$ are highlighted. Additionally, the colour indicates the tension in the beam, either positive (aft or trailing side) or negative valued (fore or leading side). Such a tension distribution is possible when the dimensionless gravity-to-stiffness ratio and inclination angle are large enough to cause compression due to drag anisotropy.}}
    \label{fig:loop_render}
\end{figure}

Uniform drag would not lead to any compression, so even for a qualitative explanation of hydrodynamic buckling it is necessary to include the dependence of the drag force density on the location on the loop. The intuition that the drag is larger in the areas where the filament is perpendicular to the flow is exemplified by a known result of local SBT: that the ratio of the drag coefficients of a slender body in directions parallel and perpendicular to the centreline of the body is $1/2$ \citep[p. 79]{Guazzelli:2012} up to $O(\epsilon)$, where $\epsilon \ll 1$ is the \rw{slenderness parameter of the body, or the filament aspect ratio}. Consider now a sedimenting loop of circular shape, as in figure~\ref{fig:loop_render}. The local gravitational force density is uniform on the circle, but we expect the drag forces to be higher at the top and the bottom in comparison to the sides, which results in the front of the circle being hydrodynamically compressed and its back being stretched. The compression can be destabilising, leading to a spontaneous shape change at the \rw{fore} side of the loop \rw{-- in sedimentation this is the lowest point of the loop --}  which can then lead to reorientation and further deformation. For stiff loops, the global balance of hydrodynamic, elastic, and gravitational forces suggests that the hydrodynamic forces are proportional to the total weight of the loop only via a geometric, dimensionless factor. \rw{Because the flow field around stable configurations of flexible loops has to be identical to that around stiff loops of the same shape, we expect this scaling to hold for flexible loops as well}. \rw{Recalling the stability threshold form to be $EI/FL^2$, and with the force scaling as mass times gravity}, this reasoning finally hints at $EI / ((\rho_L g L ) L^2)$ being the dimensionless number governing this setup, similarly to simpler buckling examples  (here, $\rho_L$ is the fibre linear density corrected for buoyancy).

\section{Local slender-body equations}

We model the fibre as a slender elastic beam of length $L$ in a viscous fluid. To account for hydrodynamic interactions, we use a local slender-body theory which is a far-field approximation of the Stokes flow due to an obstacle with a very small aspect ratio \citep{Batchelor:1970,Cox:1970,Johnson:1980}, allowing for shape parametrisation using the centreline position $\mvec{x}(s)$, where $s \in [0,L]$ is the arc length.

The Stokes approximation is valid if two dimensionless constants are very small: the Reynolds and Stokes numbers, \rw{measuring the relative importance of viscous to advective and viscous to inertial terms in the Navier-Stokes equation, respectively}. In this case, we neglect the inertial and time-dependent terms in the Navier-Stokes equations, and arrive at the Stokes equations describing the flow field $\mvec{u}$ of an incompressible fluid with viscosity $\mu$ under external body force density $\mvec{g}$

\begin{eqnarray}
\mu \nabla^2 \mvec{u} &=& \nabla p - \mvec{g}, \\
\nabla \cdot \mvec{u} &=& 0,
\end{eqnarray}
where $p$ denotes pressure. These equations are linear and thus admit the Green's fundamental solution, also called the Stokeslet, which reads

\begin{equation}
    \mvec{u}_\text{S}(\mvec{r}) = \frac{1}{8\pi\eta r}\left(\mathbb{I} + \frac{\mvec{r}\mvec{r}}{r^2} \right)\cdot\mvec{f},
\end{equation}
where $\mathbb{I}$ denotes a unit tensor and $r=|\mvec{r}|$. The Stokeslet is associated with a point force $\mvec{f}$ acting on the fluid at the origin. Notably, its derivatives are also solutions to the Stokes equations. One of particular importance is the Stokes-doublet, which has a dipolar flow character~\citep{Blake1974} and decays faster ($\sim 1/r^2$) than the Stokeslet solution.

Slender-body theory solves for the flow around a slender object of radius $r$ and length $L$, with a typical aspect ratio \rw{(slenderness parameter)} $\epsilon = r/L\ll 1$, by approximating the force density on its surface by a distribution of Stokeslets and Stokes-doublets along the centreline.

This is motivated by the idea that distributing the singularities should be sufficient to model the flow at distances large in comparison with the typical radius of the filament. Matching the 'inner' expansion of the flow field with the 'outer' flow produced by the body as a whole, and taking into account the boundary conditions on the surface of the rod, allows for expressing the centreline velocity of the filament, $\mvec{u}(s)$, in terms of two linear operators $\boldsymbol{\Lambda}$ and $\boldsymbol{{K}}$ acting on the force density applied to the filament as

\begin{equation}
    \mvec{u}(s) = - \boldsymbol{\Lambda}[\mvec{f}](s) - \boldsymbol{{K}}[\mvec{f}](s).
\end{equation}
The operators take the form

\begin{eqnarray}
    \boldsymbol{\Lambda}[\mvec{f}](s) &=& \left[(c+1)\mathbb{I} + (c-3)\pd_s\mvec{x}\,\pd_s\mvec{x}\right]\cdot \mvec{f}(s), \\
    \boldsymbol{{K}}[\mvec{f}](s) &=& \int_0^L \left( \frac{\mathbb{I} + \hat{\mvec{R}}(s,s')\hat{\mvec{R}}(s,s')}{|\mvec{R}(s,s')|} \cdot\mvec{f}(s') - \frac{\mathbb{I}+\pd_s\mvec{x}\,\pd_s\mvec{x}}{|s-s'|} \cdot\mvec{f}(s) \right) \textnormal{ds}', 
\end{eqnarray}
where $\mvec{R}(s,s') = \mvec{x}(s)-\mvec{x}(s')$, $\hat{\mvec{R}}(s,s')=\mvec{R}(s,s')/|\mvec{R}(s,s')|$, and $c = -2\log(\epsilon)$ being a function of the slenderness parameter. This method was initially developed by \citet{Batchelor:1970} and later improved by \citet{Cox:1970,Keller:1976}, and \citet{Johnson:1980}. The non-local contribution ($\boldsymbol{K}[\mvec{f}]$ together with the $c$-independent part of $\boldsymbol{\Lambda}[\mvec{f}]$) vanishes with comparison to the local one at a rate $o(1/\log(\epsilon))$. \rw{In this contribution, we shall take advantage of this asymptotic behaviour by neglecting $\boldsymbol{K}[\mvec{f}]$ entirely}. For slender fibres, neglecting the non-local term is a great simplification towards an analytical treatment of the resulting equations and leads to the local slender-body theory, also known as resistive-force theory (RFT) of \citet{Gray:1955}, in which the local velocity (in a quiescent fluid) is related to the local hydrodynamic drag force on the filament, $\mvec{f}_\text{h}$,  by

\begin{equation}
\mvec{u}(s) = -\frac{c}{8 \pi \mu}\left(\mathbb{I} + {\pd_s\mvec{x}}\,{\pd_s\mvec{x}}\right)\cdot\mvec{f}_\text{h}(s). \label{eqn:local_sbt}
\end{equation}
\rw{More recently, RFT has been increasingly popular as a modelling technique in biological fluid dynamics and the analysis of the motion of slender filaments, leading, e.g., to a general qualitative agreement with experimental observations of deforming flagella \citep{lauga_eloy_2013}, or giving insights into the buckling \citep{decanio} and swirling instabilities \citep{Stein2020} in the microtubule cytoskeleton. In fact, high-precision tracking of swimming  sperm revealed that RFT can quantitatively predict the complex trajectory of a sperm cell \citep{Friedrich1226}. Even in the case when the slender filaments come close together, RFT has proven to be useful in predicting their bundling behaviour \citep{Man_2016}.} In Stokes flow, the forces acting on a suspended body balance out to zero. In our case, this involves elastic forces, gravity, and hydrodynamic drag on the filament, so the no-net-force condition can be written as

\begin{equation}
    \mvec{f}_\text{el} + \mvec{f}_\text{g} +\mvec{f}_\text{h} = 0.\label{forcebalance}
\end{equation}
We model the elastic forces $\mvec{f}_\text{el}$ according to the Euler-Bernoulli beam theory which takes into account only the local curvature of the filament and the longitudinal tension \citep{Bernoulli:1742, Tornberg:2004}.  The elastic force density is then given by

\begin{equation}\label{elasticforce}
\mvec{f}_\textnormal{el} = EI\,\pd_{ssss}\mvec{x} - \pd_s (T(s) \pd_s\mvec{x}),
\end{equation}
where $EI$ is the flexural (bending) rigidity and $T(s)$ is the tension of the filament. The second term imposes a constant length of the filament with $T$ acting as a Lagrange multiplier.

The tension is determined by the \rw{inextensibility} equation $|\pd_s\mvec{x}| = 1$, which can be rewritten as a condition on the filament velocity by taking the time derivative

\begin{equation}
    0 = \frac{1}{2}\pd_t | \pd_s\mvec{x} |^2 = \pd_{st}\mvec{x} \cdot \pd_s\mvec{x} = \pd_s\mvec{u} \cdot \pd_s\mvec{x}, \label{eqn:inext}
\end{equation}
and noting that it is satisfied initially. One problem that arises due to this treatment is the lack of a correcting mechanism in cases when the length changes slightly due to numerical errors. We implement the solution proposed by \citet{Tornberg:2004} by introducing a numerical stabilisation term, recasting eq. \eqref{eqn:inext} as

\begin{equation} \label{eq:inext}
    0 = \pd_s\mvec{x} \cdot \pd_s\mvec{u}  - w (1 - \pd_s\mvec{x} \cdot \pd_s\mvec{x}),
\end{equation}
with $w$ controlling the absolute extension penalty.

The force balance condition, eq.~\eqref{forcebalance}, governs the dynamics. The hydrodynamic force density on the filament is determined by the sum of the gravitational and elastic forces (as in eq. \eqref{elasticforce}), which is then used to compute the velocity of the filament centreline via eq.~\eqref{eqn:local_sbt}. Equation~\eqref{eqn:inext} closes the system by imposing the filament \rw{inextensibility}. We now rescale these equations to arrive at a dimensionless system. Firstly, we choose the dimensionless arc length to have a domain $s \in [0, 2\pi]$ for additional convenience when expanding in Fourier series. This results in the characteristic length $L / 2\pi$ scaling for $\mvec{x}$. Secondly, we introduce $L^3 / (8 \pi^3 EI)$ as the force scale. Finally, we rescale the time by choosing $2 \mu L^3 / (\pi EI c)$ as our velocity scale. This leads to equations in the form
\begin{eqnarray}
    \mvec{f} &=& -\pd_{ssss}\mvec{x} + (\pd_s\mvec{x} \pd_s + \pd_{ss}\mvec{x}) T(s) - g_0 \mvec{\hat{e}_z}, \label{eqn:fp}\\
    \mvec{u} &=& \mvec{f} + (\pd_s\mvec{x} \cdot \mvec{f}) \pd_s\mvec{x} \label{eqn:up},\\
    0 &=& \pd_s\mvec{x} \cdot \pd_s\mvec{u}  - w (1 - \pd_s\mvec{x} \cdot \pd_s\mvec{x}), \label{eqn:ep}
\end{eqnarray}
with a single dimensionless parameter $g_0 = L^3 g \rho_L / (8 \pi^3 EI)$, {analogous to (the inverse of) that used by \citet{Gruziel:2019} for the bead-spring model}. Here ${\rho_L =  \pi r^2 (\rho_{\textnormal{beam}} - \rho_{\textnormal{fluid}})}$ is the mass per unit length of the fibre corrected for buoyancy, with $\rho_\text{beam}$ and $\rho_\text{fluid}$ being the densities of the beam material and the fluid, respectively. We note that $w$ is merely a numerical stabilisation constant having no influence on the solutions under exact evolution.

\section{Linear stability analysis of the planar circle solution}\label{sec:stab}

\begin{figure}
    \centering
    \includegraphics[width=0.5\linewidth]{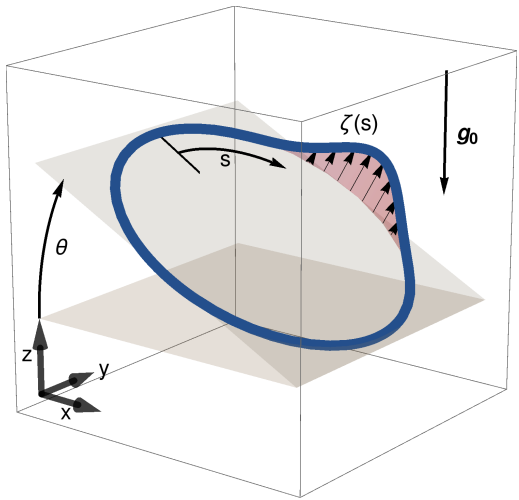}
    \caption{\rw{Diagram of the considered initial perturbation $\zeta(s)$. Presented linear stability analysis focuses only on the perturbation in the direction perpendicular to the plane in which the unperturbed solution lies.}}
    \label{fig:loop_render2}
\end{figure}

We now use the equations of motion to study the sedimentation dynamics of looped filaments with the initial condition that they are perfectly circular and inclined at an angle $\theta_0$ to the horizontal plane, as in fig.~\ref{fig:loop_render}. A rigid circle solution (including full hydrodynamics with non-local terms) was already known previously \citep{Tchen:1954,johnson_wu_1979,majumdar1977} and serves as the starting point for our stability analysis.

Equations~\eqref{eqn:fp}--\eqref{eqn:ep} admit a translating ($\pd_s \mvec{u} = \mvec{0}$) solution with a single parameter $\theta_0$ as
\begin{eqnarray}
\mvec{x}_0(s,t) &=& [\sin s, \cos \theta_0 \cos s, \sin \theta_0 \cos s] + \mvec{u}_0 t,\\
T_0(s) &=& 1 + \frac{g_0 \sin \theta_0}{3} \cos s,\\
\mvec{u}_0 &=& \left[0, \frac{g_0 \sin 2 \theta_0}{6} , \frac{ g_0 (7 - \cos 2 \theta_0)}{6}\right]. \label{eqn:sedivel}
\end{eqnarray}
Note that for $g_0 (\sin \theta_0) / 3 > 1$ there appears an area of negative tension (compression) in the beam, also the tension is largest on the aft side of the loop, explaining the observations of \citet{Alizadehheidari:2015} that \rw{such flexible loops of DNA tend to break near \rw{fore} or aft  more frequently than in between -- we propose that this effect is compounded by the higher curvature as noted in the mentioned work}.

Similarly to a sedimenting slender rod, this solution exhibits a lateral drift due to the friction anisotropy. The maximal settling angle (the angle between the sedimentation velocity and gravity) is $\gamma_{max} = \tan^{-1}(1/(4 \sqrt{3})) \approx 8.2^{\circ}$ (as compared to $\gamma_{max} \approx 19.47^{\circ}$ for a thin rod~\citep[p. 83]{Guazzelli:2012}). It is relatively small, because some parts of the circle contribute to the downwards force, while not contributing to the sideways force at any chosen angle.

We perturb the solution~\eqref{eqn:sedivel} by taking

\begin{equation}
\mvec{x}(s) = \mvec{x}_0 + \mvec{\tilde{x}} + O(\mvec{\tilde{x}}^2),
\end{equation}
where a tilde over a symbol denotes the perturbation function.
A general form of the perturbation turns out to be analytically intractable due to the complexity of the coupled equations for tension perturbation, so further simplifying assumptions are necessary. Here, we consider specific perturbations in the direction perpendicular to the circle's plane, so that $\mvec{\tilde{x}}$ is of the form

\begin{equation}
\mvec{\tilde{x}} = \zeta(s) [0, -\sin \theta_0, \cos \theta_0].
\end{equation}
\rw{The presented method gives rise to two problems when trying to expand to in-plane perturbations. Firstly, taking a dot product with an in-plane vector instead of a normal vector leads to significantly more complex equations. The second complication is that in-plane perturbations are inherently two-dimensional and cannot be described by just a single scalar function. Choosing only specific normal perturbation is justified by the intuitive insight that comes from the tractable form of the resulting linear stability analysis problem.}

Assuming that the associated perturbation of tension $\tilde{T}$ is $O(\zeta)$, we neglect quadratic and higher order terms in $\zeta$. Then the force $\mvec{f}$ in the perturbed system is of the     form $\mvec{f} = \mvec{f}_0 + \mvec{\tilde{f}}$ where

\newcommand{\xt}{\mvec{\tilde{x}}}
\newcommand{\xz}{\mvec{x}_0}
\begin{eqnarray}
    \mvec{f}_0 &=& \frac{ g_0 \sin \theta_0}{3} \, \left[\sin 2s,\, \cos 2s \cos \theta ,\, \frac{3}{\sin \theta} + \cos 2s \cos \theta\right],\\
    \mvec{\tilde{f}} &=& - \pd_{ssss} \xt + \pd_s T_0 \, \pd_s \xt + \pd_s \tilde{T} \, \pd_s \xz + \tilde{T} \, \pd_{ss} \xz + T_0 \, \pd_{ss} \xt.
\end{eqnarray}
Finally, we get a linear resulting perturbation to the velocity $\mvec{u} = \mvec{u}_0 + \mvec{\tilde{u}}$ with

\begin{equation}
\mvec{\tilde{u}} = \mvec{\tilde{f}} + (\mvec{\tilde{f}} \cdot \pd_s \xz) \pd_s \xz + (\mvec{f}_0 \cdot \pd_s \xt) \pd_s \xz + (\mvec{f}_0 \cdot \pd_s \xz) \pd_s \xt.
\end{equation}
This can be put into the \rw{inextensibility} condition $\pd_s\mvec{u} \cdot \pd_s\mvec{x} = 0$. With $\pd_s \mvec{u}_0=\mvec{0}$ and ${\pd_s \mvec{\tilde{u}} \cdot \pd \tilde{\mvec{x}}}$ being a second order term, the inextensibility equation for the perturbed shape is simply

\begin{equation}
0 = \pd_s \mvec{\tilde{u}} \cdot \pd_s \xz.
\end{equation}
It could be in principle solved for the tension perturbation. Instead, a more convenient way of proceeding is to note that

\begin{equation}
\begin{split}
\pd_t \zeta = \mvec{\tilde{u}} \cdot [0, -\sin \theta_0, \cos \theta_0] = \pd_{ss}\zeta + \frac{g_0 \sin \theta_0}{3} \left[(\pd_s\zeta) \sin s +(\pd_{ss}\zeta) \cos s \right] - \pd_{ssss}\zeta = \mathcal{L}[\zeta]\label{eqn:stab}
\end{split},
\end{equation}
which has all the information needed to analyse the  evolution of perturbation in the direction of the initial perturbation. The perturbation dynamics are now governed by a single parameter only
\begin{equation}
    q = \frac{1}{3} g_0 \sin \theta_0.
\end{equation} Equation \eqref{eqn:stab} can be rewritten as a diffusion-like equation of the form

\begin{equation}
\pd_t \zeta = - \pd_{ssss}\zeta + \partial_s(T_0 \partial_s\zeta) \label{eqn:difu}.
\end{equation}
This highlights the essential role of negative tension in the development of shape instability, which takes the role of the diffusion coefficient in the equation \eqref{eqn:difu} and the only other term is $\pd_{ssss}\zeta$, which has an additional stabilising effect.

To determine the stability of the linear PDE \eqref{eqn:difu}, we examine the eigenvalues of the linear operator $\mathcal{L}$ on the right-hand side. The value of the initial tension $T_0(s) = 1 + q \cos s$ gives rise to a simple analytical form of this operator

\begin{equation}
\mathcal{L} = - \pd_{ssss} + \pd_{ss} + q ( (\sin s) \,\pd_s + (\cos s) \,\pd_{ss}).
\end{equation}

The periodicity of $\zeta$ can be enforced by analysing $\mathcal{L}$ action on $\{\sin ks,\cos ks\}$ basis on $L^2(S^1)$. We note that

\begin{equation}
\begin{split}
\mathcal{L}[\sin ks] = 
-(k^2+k^4)\sin ks  + \frac{1}{2}kq(k+1)\sin(k-1)s + \frac{1}{2}kq(k-1)\sin(k+1)s.
\end{split}
\end{equation}
and similarly for $\cos ks$.  Therefore, to find the eigenvalues of $\mathcal{L}$, it is sufficient to consider linear combinations of the sine and cosine parts of the Fourier expansion separately, as $\mathcal{L}$ maps the span of either one to itself. Moreover, on each of the subspaces, the restricted maps are the same and thus have identical eigenvalues.

The operator $\mathcal{L}$ on $\operatorname{span}[\sin(kx)]$ has the following matrix representation

\begin{equation}
\mathcal{L} = \begin{bmatrix}
-2 &-3q&   &    &      &&\phantom{\ddots}\\
 0 &-20&-6q&    &      &&\phantom{\ddots}\\
   &-q &-90&-10q&      &&\phantom{\ddots}\\
   &   &-3q&-272&      &&\phantom{\ddots}\\
   &   &   &-6q &\ddots& -qk(k+1) /2&\phantom{\ddots}\\
   &   &   &    &      & -k^2-k^4   &\phantom{\ddots}\\
\phantom{\ddots} & \phantom{\ddots}  & \phantom{\ddots}  & \phantom{\ddots} &  \phantom{\ddots} & -qk(k-1)/2 &\ddots\\
\end{bmatrix}.
\end{equation}

\begin{figure}
	\centering
	\includegraphics[width=0.9 \linewidth]{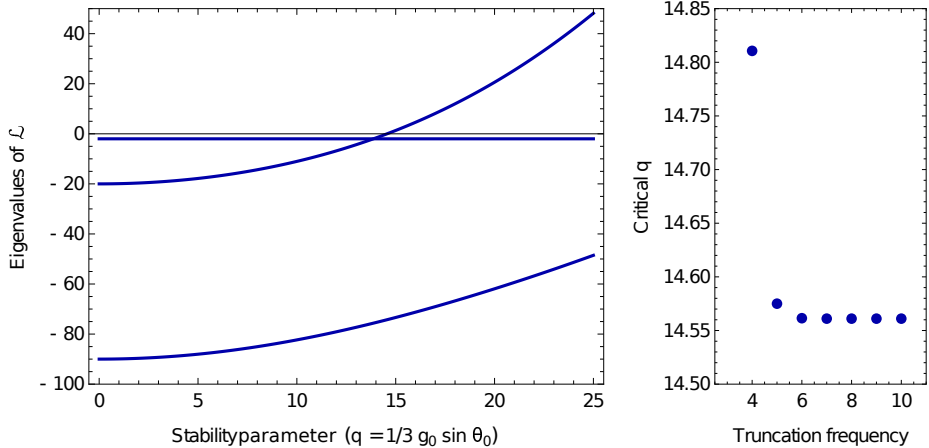}
	\caption{Left: Largest eigenvalues of the linear operator $\mathcal{L}$ depending on the stability parameter $q$ (dark blue lines) describing the linearised stability problem. At $q$ around $14.5$ the largest eigenvalue crosses zero, which corresponds to the appearance of an unstable solution of the time-dependent equation. Right: The critical value of $q$ computed for various values of the truncation frequency $n$. For $n<4$ the behaviour of $\mathcal{L}$ is completely different, but for $n\geq 4$ the critical value of the stability parameter changes by a very small fraction. This is possible because the most unstable mode is dominated by low-frequency oscillations.}
	\label{fig:eigen}
\end{figure}

We can obtain approximations to $\mathcal{L}$ by truncating at a desired $n$. For a given $n$ the condition that $q$ is critical translates to $\mathcal{L}$ having one eigenvalue equal to zero, which can be expressed as $\det \mathcal{L} = 0$, which is a polynomial equation in $q$. Such equations have fast numerical solvers allowing for computation of the critical value with high accuracy. We examined this for $n \in (1, 2, \dots 60)$ to verify that the highest eigenvalue of $\mathcal{L}$ was determined with satisfactory precision -- the convergence is extremely fast (at least exponential), as illustrated in figure~\ref{fig:eigen}. We find that the critical value of $q = g_0 (\sin\theta_0)/3$ is $14.56105439107$. Above this critical value the largest eigenvalue is positive, as illustrated.

\section{Numerical method}

In order to verify the predictions of the theoretical model and the simplified linear stability analysis, we solve the equations of motion numerically. Because all the functions characterising the elastic loop are periodic, we represent them in the form of (truncated) Fourier series. More precisely, an approximation to a function $f(s)$ is numerically represented by a complex valued $2n$-dimensional vector $f_\alpha$, such that:

\begin{equation}
f(s) \approx \sum_{\alpha = -n}^{n} f_\alpha \exp(i \alpha s).
\end{equation}

For a smooth $f$, such series converge exponentially. We simulate the equations of motion by computing truncated series approximations to the position $\mvec{x}$, velocity $\mvec{u}$, and tension $T$ up to a fixed order $n$.

The governing equations can be coded as three affine maps (of type $\mvec{x} \mapsto \mathsfbi{A} \mvec{x} + \mvec{b}$) as follows:

\rw{
\begin{equation}
T \xrightarrow[(1)]{\textnormal{Euler-Bernoulli equation}} \ 
\mvec{f} \xrightarrow[(2)]{\textnormal{local SBT}} \
\mvec{u} \xrightarrow[(3)]{\textnormal{inextensibility equation}} \ 
\varepsilon, \label{eqn:maps_sim}
\end{equation}
}
\rw{These correspond to the Euler-Bernoulli theory (map $(1)$ -- equation \eqref{elasticforce}) allowing for computation of the force density given the tension $T$, local SBT mapping force density $\mvec{f}$ to the local velocity $u$ (map $(2)$ -- equation \eqref{eqn:local_sbt}), and the inextensibility equation mapping the local velocity to local length creation, the error term $\varepsilon$, which we try to minimise in the simulation (map $(3)$ -- equation \eqref{eq:inext})}. Maps $(1),(2)$ are approximated by matrix equations of dimension corresponding to the truncation order $n$. This is chosen in order to keep $\mvec{x}$ and $\mvec{u}$ expanded to the same order. Nevertheless, it is essential to compute the inextensibility equation map $(3)$ including higher order ($2n$) terms. \rw{These three maps are combined to obtain a relationship between the coefficients of the Fourier expansion of the tension distribution $T$ and the local filament length creation $\varepsilon$, which should be as close to zero as possible. Because there are more terms in the expansion of the local length creation than in the tension distribution we can only attempt to make them as close to zero as possible.} The combined affine map from the tension $T$ to the error term $\varepsilon$ therefore induces an overdetermined system of linear equations. These are solved for $T$ by $L^2$ error minimisation, which is the same as solving the ordinary least squares (OLS) problem (Fourier basis is orthonormal) with an additional restriction that $T(s)$ is real-valued. \rw{It might be tempting to simplify this procedure by using the same truncation order on the error terms as on the tension expansion, leading to an exact solution for $T$ instead of an optimisation problem, but} if we leave out the highly oscillatory terms in the error map $(3)$, then the solver is oblivious to the filament length increase due to oscillations with frequencies higher than $n/2$ \rw{in the tension expansion}, resulting in an exponential explosion of high frequency vibrations. \rw{Such behaviour comes from the terms in the equations where two functions are multiplied (such as $\pd_s T\, \pd_s \mvec{x}$); there two terms of a given wavenumber can combine to one term with double the wavenumber.} Additional care needs to be taken to ensure that the trajectories remain real-valued (the error can accumulate in the complex-valued $\mvec{x}$ Fourier expansions). This was achieved by projecting the solution onto the allowed subspace at each time step.

Equations \eqref{eqn:maps_sim} for the tension $T$ are solved at each evaluation of $\mvec{u}$, and this value of tension is used to compute the velocity in an explicit integration scheme with a variable timestep of the Runge-Kutta-Feldberg (fifth order) method. This algorithm, however, is at best $O(n^3)$ in the truncation frequency (because the OLS minimizer is $O(n^3)$) and in reality even slower, as more degrees of freedom necessitate a decrease of the time step. For $n=6$, our implementation was running at a speed of $40$ dimensionless time units per hour for typical values of parameters on one thread of a typical 2.5 GHz processor; for $n=8$, the speed decreased to $5$ dimensionless time units per hour (giving a very crude estimate for complexity of about $O(n^7)$). This makes investigations of large values of $n$ impractical. For our calculations, we choose $n=6$. We discuss this choice further in Sec. \ref{sec:truncation_order}. Most of the simulations were run with the help of GNU Parallel software \citep{Tange2011a}.

\section{Sedimentation modes}

\begin{figure}
	\centering
	\includegraphics[width=\linewidth]{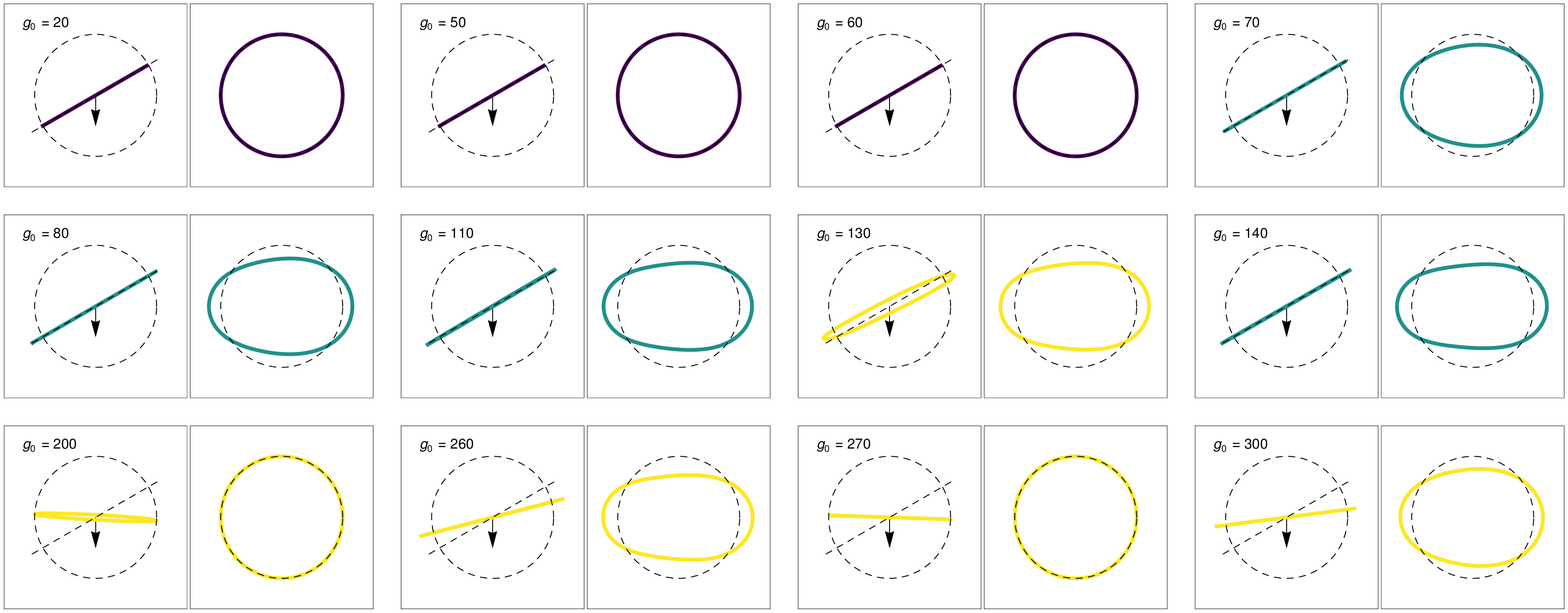}%
	\caption{Selection of terminal shapes for varying values of forcing \rw{(inverse stiffness)} $g_0$ with an initial angle $\theta_0 = 30 ^\circ$ and $n = 6$ coloured by trajectory type: stable (dark purple), \rw{in-plane} dynamics (green-blue), 3D dynamics (yellow). Left panels show side view (sedimentation downwards) with a unit circle and initial tilt plane \rw{marked by dashed lines}; black arrows indicate the direction of gravity. \rw{When the terminal shape aspect ratio is 1, regardless of the terminal inclination angle, the projection fits inside the unit circle}. Right panels show the shape within the final sedimentation plane \rw{aligned with the principal axes of the loop. Right-left axis on the graphics thus corresponds to the fore-aft direction in the terminal configuration}. Dashed unit circle is plotted for reference. \rw{Note that for sufficiently large values of $g_0$ (highly flexible filaments) the terminal tilt angle changes erratically with initial conditions.}}%
	\label{fig:galery_t_30}
\end{figure}

\begin{figure}
	\centering
	\includegraphics[width=0.85\linewidth]{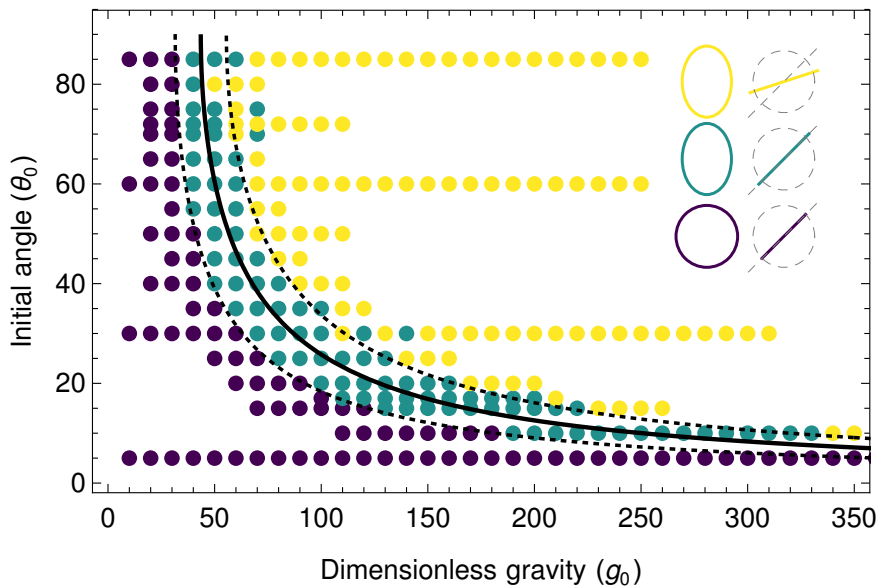}%
	\caption{Sedimentation regimes showing \rw{stable translation} (dark purple), \rw{in-plane} dynamics (green-blue), and 3D dynamics (yellow), observed in simulations for truncation frequency $n=6$. Solid line represents the linear stability analysis prediction ($q \approx 14.5$); dashed lines are empirical \rw{in-plane} dynamics boundaries ($q \approx 10.5$ and $q \approx 18.5$) presented as eye-guides.}%
	\label{fig:phase_n_6}
\end{figure}

\begin{figure}
	\centering
	\includegraphics[width=\linewidth]{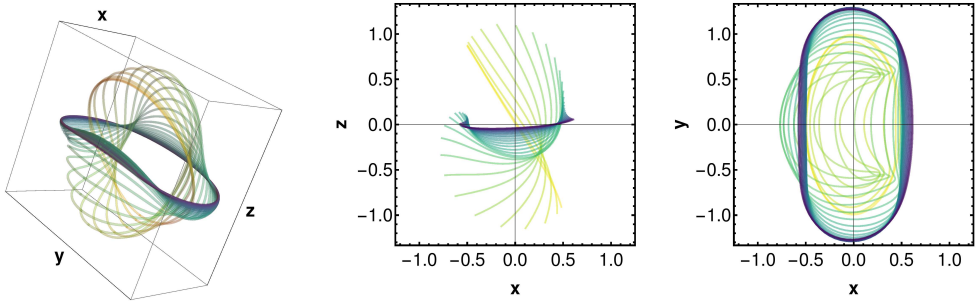}%
	\caption{Sample of 3D time evolution dynamics. 3D render (left), side view (centre), top view (right). Images show shape evolution when starting in an unstable equilibrium.  Snapshots are taken at regular time intervals with changing colour: initial light yellow to final dark purple. The loop starts at initial angle $\theta_0 = 60^\circ$ subject to gravity force of dimensionless value $g_0 = 150$. The \rw{fore} side of the loop folds upwards leading to the formation of two lobes and an ellipsoidal shape ( best visible in top view). At long times, the loop converges to a \rw{near-horizontal} plane to finally (after very slow dynamics) relax to a perfect circle (not shown in the figure).}%
	\label{fig:3ddyn}
\end{figure}

\begin{figure}
	\centering
	\includegraphics[width=\linewidth]{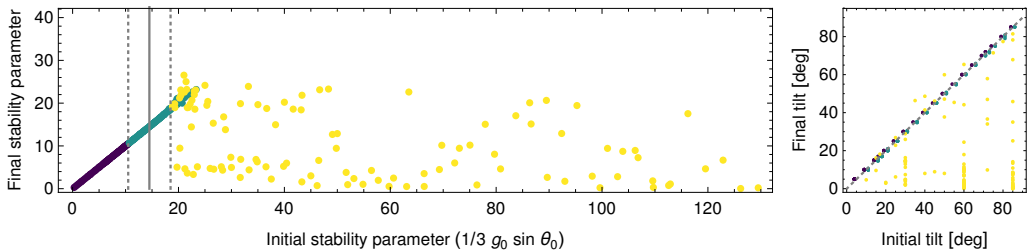}%
	\caption{Left: A comparison between the initial and the final value of the stability parameter in all simulations. Linear stability analysis prediction (solid line at $q \approx  14.5$) is marked together with estimated \rw{in-plane} regime boundaries of $q \approx 10.5$ and $q \approx 18.5$ (dashed lines). The points are coloured by the type of trajectory: stable (dark purple), \rw{in-plane} dynamics (green-blue), 3D dynamics (yellow). Right: A comparison between the initial and the final value of tilt angle. The stable (darkest) and the \rw{in-plane} dynamics (medium) points were shifted $1^\circ$ to the left and right respectively, for clarity. The unordered scatter of the 3D dynamics points (yellow) shows that the final tilt is difficult to predict, but in the vast majority of cases it is smaller than the initial tilt angle.}%
	\label{fig:stability_parameter}
\end{figure}

\begin{figure}
	\centering
	\includegraphics[width=\linewidth]{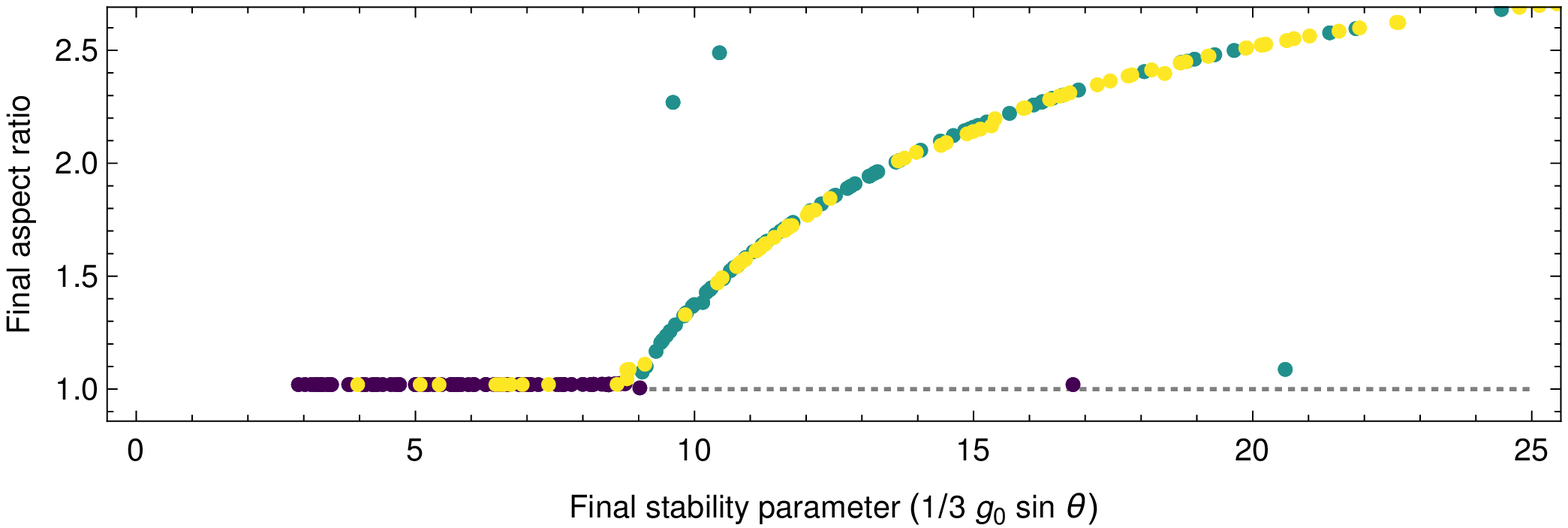}%
	\caption{\rw{Aspect ratio of the terminal centreline shape plotted against the terminal value of stability parameter $g_0 (\sin\theta) /3$. Large aspect ratios refer to highly elongated loops}. A functional relationship confirms that the \rw{in-plane} dynamics depend only on the stability parameter. The final shape is determined by the final sedimentation angle (and vice versa). Regime boundary between stable and \rw{in-plane} dynamics is clearly visible on this graph as a point where medium shade points reach aspect ratio $=1$ line. \rw{When initial stability parameter is greater than $q^\ast\approx 9$, loops move from stable to \rw{in-plane} dynamics regime.} Note that for the stable and \rw{in-plane} dynamics trajectories (darkest and medium points) the initial and the final tilt are the same, but for 3D dynamics the terminal tilt angle is smaller than the initial (explaining lightest points with aspect ratios of unity). \rw{Regardless of initial configuration and parameters, it is sufficient for the terminal stability parameter to be smaller than $q^\ast$ to ensure the terminal configuration taking the shape of a circle. Dashed line shows the unstable branch of the solutions.}}%
	\label{fig:aspet_v_stability}
\end{figure}
%
%
{Motivated by the linear stability analysis performed above for circular loops, we now explore numerically the evolution of elastic rings starting at arbitrary inclination angles for a range of the elastohydrodynamic parameter $g_0$. We thus choose similar initial conditions to those used by \citet{Gruziel:2019}.} In the simulations, we observe three distinct sedimentation regimes  \rw{(corresponding to terminal shapes)} depending on the stiffness of the loop. They are demonstrated in figure~\ref{fig:galery_t_30} with top and \rw{in-plane} views of the terminal sedimentation shapes. For very stiff loops, characterised by low values of $g_0$, marked in purple in the figure and referred to as \emph{stable}, we observe stable sedimentation of a circular shape. The dynamics are then given by the translating solution of eq.~\eqref{eqn:sedivel} -- \rw{we see no change in the shape or the sedimentation angle}. When the elasticity of the loop is increased, we observe a different terminal regime referred to as \emph{in-plane dynamics}, where the shape of the loop evolves in a two-dimensional plane defined by the initial angle (cases marked in green-blue) -- \rw{we only see changes of the shape, while the inclination angle of the loop remains unchanged}. The circular state is then unstable and evolves into a prolate loop. When the stiffness is reduced further, for high values of $g_0$, we observe three-dimensional evolution (marked in yellow and referred to as \emph{3D dynamics}) with a change in the \rw{angle of the terminal sedimentation plane} with respect to the initial inclination. \rw{In the \emph{3D dynamics} regime the loop leaves the initial plane essentially immediately – it does not go through dynamics similar to the \rw{in-plane} regime. Deviations from a circular shape are necessary but not sufficient for a change in the inclination angle, as exemplified by the existence of the  \emph{in-plane dynamics} regime. For the \emph{3D dynamics} regime in each simulation we observe \rw{in-plane} and out-of-plane perturbations appearing spontaneously every time, and the inclination angle always changes.}

In order to test the stability criterion derived in Sec.~\ref{sec:stab}, we plot the observed sedimentation modes depending on the dimensionless gravity $g_0$ and the initial sedimentation angle $\theta_0$ in fig.~\ref{fig:phase_n_6}. Clearly, the stability criterion with $q\approx14.5$ (solid line) divides the regions of absolute stability (with purple markers) and full 3D dynamics (yellow markers), with the planar evolution states in between. All instances of stable behaviour are inside the predicted stability region. \rw{The initially unstable behaviour – yellow points to the right of the stability curve – involves a complex transient evolution that finally settles on a stable configuration at a smaller inclination angle}. Even though the final shape in fig.~\ref{fig:galery_t_30} is planar (prolate or circular), an example of the full shape transition in figure~\ref{fig:3ddyn} shows significant bending with a complete deviation from the initial plane and the establishment of a new ellipsoidal shape in a different plane, essentially always at an angle smaller than the initial $\theta_0$, and followed by a relaxation to the final shape. In most cases, the trajectory consists of three phases. Firstly, the loop folds in half starting with the \rw{fore} side of the loop \rw{falling faster than the centre of mass and immediately after} being 'blown' backwards by the drag force. \mkl{This is related to the loop deforming towards a more prolate shape. With the increasing fore-aft distance, and the centre of mass position remaining symmetrically in the middle of the loop, it is necessary for the fore side to move faster than the centre of mass during this stage. We regard this as the primary reason for the different behaviour of flexible rods and loops. In the initial stages of motion a loop extends in-plane, elongating its long axis, while a rod retains its constant length. We note, however, that this effect lasts for a very short period of time, and is present only at the early stages of the evolution.} Secondly, the two lobes formed by the fold relax towards the terminal plane. The dynamics then become very slow and the loop attains the terminal shape within the terminal plane. \rw{Notably, this behaviour is different from the case of a free-end filament in which the lowest part of the filament initially sediments slower than the centre of mass \citep{Lei:2013}, leading to a different shape evolution path}.

 Lastly, the green-blue points span across the stability boundary in fig.~\ref{fig:phase_n_6}. This type of behaviour is not taken into account by our simplistic linear stability analysis, because of the assumption of a normal direction of perturbation, while here the loop stays in the initial plane. Nevertheless, these types of dynamics are observed in the vicinity of the predicted stability boundary and can be regarded as an intermediate stage between complete instability and the complete lack of shape change. A vast majority of such trajectories are bounded by $10.5< (g_0 \sin \theta_0) /3 < 18.5$. This region closely follows the stability curve in a wide range of the control parameter values and has boundaries that appear to have a similar functional form to the analytical predictions. 
 
 To further analyse the relationship between the initial and final sedimentation plane, in figure~\ref{fig:stability_parameter} we plot the initial stability parameter $q=g_0\sin\theta_0/3$ versus the final parameter (calculated as $g_0\sin\theta/3$) for all the cases investigated. The stability threshold is again marked with a solid line, together with the empirical strip of 2D-evolving shapes \rw{between the dashed lines}. For stiff loops, at low values of $q$, the evolution does not affect the sedimentation angle, and thus we see the expected linear correlation, which persists for the semi-stable states which still remain in the initial plane. For unstable loops of high flexibility, we see no apparent correlation between the initial and the final sedimentation angle, as seen clearly in fig.~\ref{fig:stability_parameter} (right panel). \mkl{Although the transition from stable to unstable dynamics is quite accurately grasped by our estimates, we note that the detailed evolution in this mode of motion is sensitive to the initial conditions, as discussed further.}
 
 In fig.~\ref{fig:aspet_v_stability}, we explore the deviation from a circular shape versus the final stability parameter. To this end, we plot the aspect ratio of the final shape for all our data, \rw{defined as the ratio of the highest to the middle eigenvalue of the spatial covariance matrix}. The absolutely stable loops remain circular, and once we enter the unstable region, in the vicinity of the stability criterion we see a continuous increase of the aspect ratio. For unstable loops undergoing the full 3D evolution, we do not see a systematic trend, but the final aspect ratios remain within the trend seen for stiffer loops, confirming that these configurations are globally stable solutions of the loop evolution equations. In addition, we see that the planar evolution is governed solely by the stability parameter $q = g_0 \sin \theta/3$, also in the cases where the final value of $\theta$ is very small and the initial 3D dynamics converge to a perfectly circular equilibrium shape, the same shape as in the case of stable sedimentation modes.
 
\subsection{Influence of truncation order}\label{sec:truncation_order}

All the presented summaries of simulations are results of a numerical scheme terminated at $t=10$ of the dimensionless time, which corresponds to the sedimentation distance of the order of $10^3$ in terms of loop radius for moderate values of stiffness, upon confirming no further shape evolution. This is long enough for all the simulated shapes to attain the final configuration. In most cases, all rapid changes in shape (each at timescales of about $0.5$) are finished when $t$ reaches about $3$, then the terminal angle is selected. A much slower relaxation of shape within the terminal plane follows with the characteristic time of $t\approx1$.

In the section above, we studied the stability of initial configurations and the attraction to a stable shape for different values of the initial angle $\theta_0$ and the truncation orders $n=6$. Numerical investigation of the eigenvalues of the truncated $\mathcal{L}$ operator shows that the stability boundary should be largely independent of the truncation order $n$. 
Even though the initial rate at which the instability develops will be independent of $n$ (in an unstable equilibrium it is the numerical noise that initiates movement), we should expect quantitatively similar behaviour for the terminal motion of the loops.

\begin{figure}
	\centering
	\includegraphics[width=0.48\linewidth]{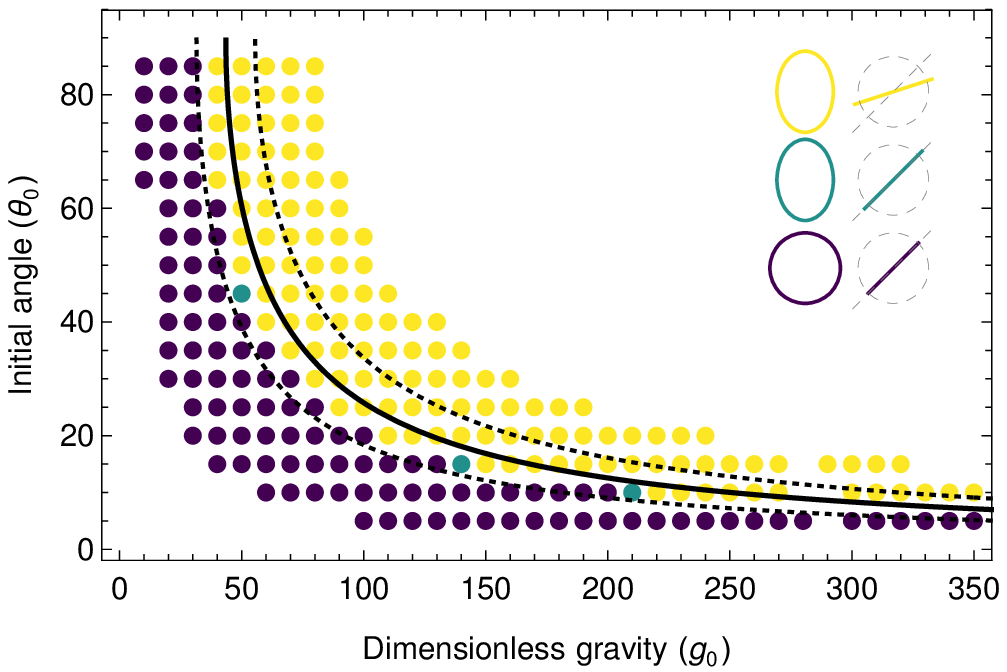}%
	\phantom{a}%
	\includegraphics[width=0.48\linewidth]{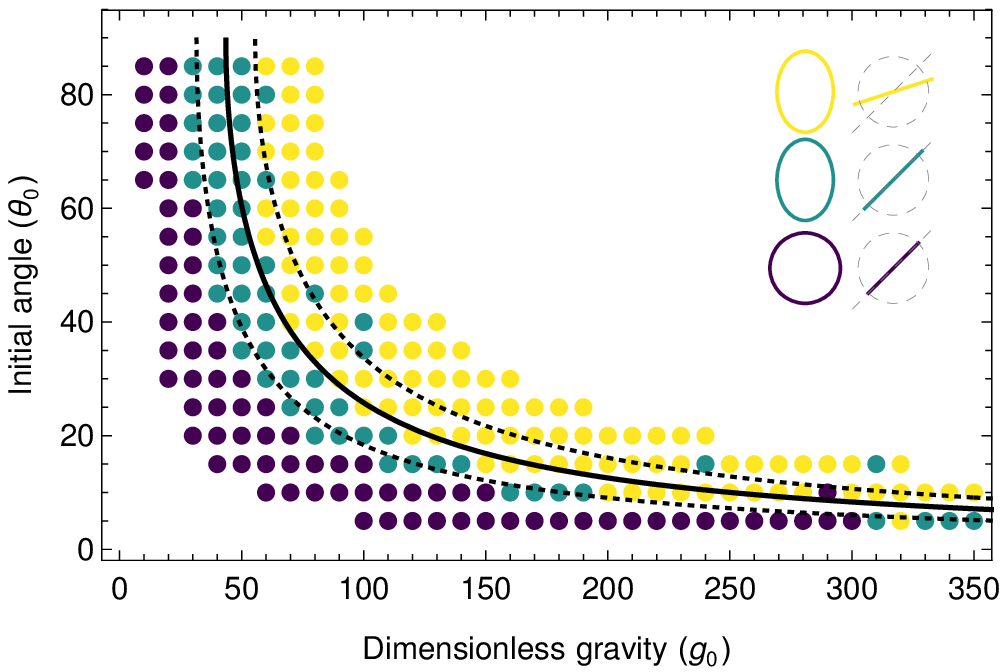}
	\caption{Comparison of  regimes of sedimentation observed in simulation for truncation frequency $n=4$ (left) and $n=8$ (right). Solid line represents the boundary of the stable region as obtained from the linear stability analysis. Dashed lines are eye-guides to facilitate the comparison of the two panels. Very small values of $n$ do not exhibit the intermediate regime of in-plane dynamics. However, the proposed stability criteria seem to hold sway.}%
	\label{fig:phase_n_4_8}
\end{figure}

In fig.~\ref{fig:phase_n_4_8}, we redraw the stability diagram, fig.~\ref{fig:phase_n_6}, for different truncation order of $n=4$ (left) and $n=8$ (right). Regardless of the initial angle and the truncation order, the stability region is correctly predicted by the simplified linear stability analysis. Moreover, when the loops are initially in an unstable position, they eventually reach the terminal angles, which are constrained by the stability region. However, the truncation order changes the exact behaviour predicted for a particular stiffness and initial sedimentation angle. Notably, for very small values of $n$ the intermediate regime of \rw{in-plane} dynamics cannot be observed. It is possible that this effect appears because the loop is unable to attain configuration close enough to the terminal shape of the \rw{in-plane} dynamics, due to the small number of degrees of freedom. For a larger number of truncation modes, we see the extent of the 2D-evolution region decrease, with more unstable states close to the stability criterion, but nevertheless the clear division between the modes remains in place, rendering the simplistic criterion to be a useful tool for assessing the loop stability.

\rw{Far from the stability boundary, the dynamics of the loop follow the qualitative conclusions derived from the analytical treatment, and the lower boundary (\emph{stable} to \emph{in-plane} dynamics) is \mkl{weakly} sensitive to the truncation frequency \mkl{-- additional simulations for $n=10$ on a restricted set of initial angles confirm that a lower stability boundary shows no further change with inclusion of higher order terms in expansions}. We note, however, that the upper boundary (\emph{in-plane} to \emph{3D} dynamics) is a fragile one, and the selection of one attractor over the other is sensitive to the details of numerical implementation. The discussed discrepancies can be of three origins: (i) change in stability due to the truncation of high frequencies, (ii) change in perturbation power spectrum due to the change of dimensionality, or (iii) change in perturbation due to the change in numerical stability of the OLS minimizer procedure. The presented analysis of truncation order in fig.~\ref{fig:eigen} gives us confidence that the high-frequency modes have negligible contributions to the stability problem due to the extremely high damping by elastic $\pd_{ssss}$ terms. Therefore, our expectation is that effects (ii) and (iii) are the primary reasons for the observed differences between the smallest and largest values of the truncation order.}

\section{Conclusions}
 In this contribution, we modelled the behaviour of  elastic loops sedimenting under gravity. To this end, we combined the local slender-body theory with the Euler-Bernoulli beam theory to develop analytical insights into the dynamics and proposed a Fourier basis expansion method for effective numerical implementation. Our approach takes advantage of the periodicity of all the relevant functions in this setting, complementing our analytical treatment.

In simulations, when starting from an inclined circle, we identified three distinct regimes of motion, depending on the relative importance of gravity and loop stiffness, combined into a single dimensionless parameter $g_0$. For stiff loops, or low values of $g_0$, we see no effect of elasticity and the loops sediment as circular rings. When increasing the softness, sedimenting loops remain oriented in the initial plane but attain elongated and slightly asymmetric shapes. For even softer filaments, the loops exhibit a transient instability, undergoing a 3D shape evolution, where the \rw{fore} edge of the loop is bent and leaves the initial plane, but the dynamics eventually settle on a planar shape at an angle different than the original. \rw{The said terminal angle is hard to predict and under fixed $g_0$ determines whether the final configuration will be circular or prolate. Small enough angles corresponding to terminal stability parameter $q=g_0 \sin\theta/3$ smaller than approximately $q^\ast = 9$ always result in an eventual circular shape.}

To explain the transition between these regimes, we propose a simple theory based on the linear stability analysis, with a further assumption that perturbation is taken in the direction perpendicular to the initial plane. \rw{This specific choice of perturbation leads to an analytical insight into the dynamics which we doubt to be possible with an arbitrary perturbation. To circumvent this difficulty, we additionally perform numerical simulations to provide a description for arbitrary initial conditions. Under the chosen simplification we identify the most important parameter controlling the motion being the initial stability parameter $q = g_0 \sin{\theta_0} /3$, and perform a near-analytic determination of the stability boundary at $q \approx 14.5$.}

The results of numerical simulations are in satisfactory agreement with the simplistic approach of the linear stability analysis, thus confirming the validity of our approach for finding the stability threshold. Both the absolutely stable and the unstable regimes fit entirely within the domains predicted by the theory. The intermediate regime of planar shape evolution appears in close proximity of the stability boundary. \rw{We conclude that such a simplified linear stability analysis is a useful tool in both the 3D dynamics case as well as for \rw{in-plane} dynamics because it correctly predicts both scaling and approximate values of the stability parameter of regime transitions}. For sufficiently stiff loops, we compare our numerical codes with the existing analytical results of \citet{johnson_wu_1979} and \citet{Tchen:1954} for stiff loops. \rw{We confirmed the expected agreement both qualitatively (translation without change of orientation) and quantitatively in terms of translation velocity asymptotics for very slender rods}. {Below the stability threshold, however, we see differences from the bead-spring model \citep{Gruziel:2019}, where sufficiently stiff sedimenting loops attained vertical or tilted oval shapes, in contrast to loops sedimenting without a change of orientation in our model.} Increasing the flexibility leads to a deviation from the initial shape, resulting in an approach to a different equilibrium circle.  \rw{Beyond the stability threshold, we also find that the details of intermittent evolution of more flexible fibres differ between the bead-spring results of \citet{Gruziel:2019} and slender-body models. This might be partly due to the lack of non-local terms in our resistive-force SBT and due to the different geometric details of both systems, i.e. a slender filament vs. a chain of beads. In particular, we note that the stable circular configuration found in the RFT approximation is no longer a solution when full hydrodynamics are included. However, the shape of the final tilt angle vs. stiffness curve is similar in the bead-spring and RFT models.}

The presented approach shows an attractive interpretation of the compression (negative tension) on the \rw{fore} side of sedimenting objects as a negative diffusion coefficient in the governing equation of the linear stability analysis. {This gives intuitive grounding to the experimental results such as those of \citet{Jay:1972} \rw{where red blood cells show a preference for horizontal sedimentation when their flexibility is increased}, or \citet{Gruziel:2018} where the preference for horizontal sedimentation was seen for knotted elastic fibres. It also provides support for the interpretation of experiments of the DNA loop rupture dynamics \citep{Alizadehheidari:2015}, in which the loops break in locations corresponding to the maximal tension in our description.}

\rw{The conclusion that vertically oriented loop configurations are forbidden due to their instability is a general physical observation applicable in similar elastohydrodynamic settings. The presented results show that observations of instability from free-end simulations \citep{Lei:2013} are applicable only to some extent: circular configuration gives rise to a tension offset which substantially improves stability in comparison to free-end configuration.}

\rw{We look forward to additional experimental verification of the conclusions of this paper, either in the micro scale with biological fibres or in macroscopic experiments such as those with knotted bead chains of \cite{Gruziel:2018}.} 

\begin{acknowledgements}
We would like to thank Maria Ekiel-Jeżewska, Magdalena Gruziel-Słomka, and Bogdan Cichocki for discussions. Science Club Fenix is acknowledged for providing computational power for simulations.

{\bf Funding} The work of ML and RW was supported by the National Science Centre of Poland grant Sonata to ML no. 2018/31/D/ST3/02408. PS was supported by the National Science Centre (Poland) under research grant 2015/19/D/ST8/03199.

{\bf Declaration of Interests} The authors report no conflict of interest.
\end{acknowledgements}



\bibliographystyle{jfm}
\bibliography{sedimenting-loops}

\end{document}